# Novel lithium-nitrogen compounds at ambient and high pressures

Yanqing Shen,[1,4,*] Artem R. Oganov,[2,3,4,5] Guangri Qian,[4] Huafeng Dong,[4] Qiang Zhu,[4] Zhongxiang Zhou[1]

1 Department of Physics, Harbin Institute of Technology, Harbin 150001, China

2 Skolkovo Institute of Science and Technology, 100 Novaya St., 143025 Skolkovo, Russia

3 Moscow Institute of Physics and Technology, 9 Institutskiy Lane, Dolgoprudny City, Moscow Region 141700, Russia

4 Department of Geosciences, Center for Materials by Design, and Institute for Advanced Computational Science, Stony Brook University, Stony Brook, New York 11794, USA

5 School of Materials Science, Northwestern Polytechnical University, Xi'an 710072, China

* Corresponding author, E-mail: shenyanqing2004@163.com

**Abstract**

Using ab initio evolutionary simulations, we predict the existence of five novel stable Li-N compounds at pressures from 0 to 100 GPa ($Li_{13}N$, $Li_5N$, $Li_3N_2$, $LiN_2$, and $LiN_5$). Structures of these compounds contain of isolated N ions, $N_2$ dimers, polyacetylene-like N chains and $N_5$ rings, respectively. The structure of $Li_{13}N$ consists of Li atoms and $Li_{12}N$ icosahedra (with N atom in the center of the $Li_{12}$ icosahedron) – such icosahedra are not described by Wade-Jemmis electron counting rules and are unique. Electronic structure of Li-N compounds is found to dramatically depend on composition and pressure, making this system ideal for studying metal-insulator transitions. For example, $LiN_3$ undergoes a sequence of pressure-induced transitions: metal-insulator-metal-insulator. This work resolves the previous controversies of theory and experiment on $Li_2N_2$.

**Introduction**

Li-N system contains two well-known compounds: lithium nitride ($Li_3N$) and lithium azide ($LiN_3$). $Li_3N$ has potential use as an electrolyte in Li-batteries and a hydrogen storage medium.[1-4] Extensive experimental and theoretical investigations show that $Li_3N$ undergoes a sequence of phase transitions with increasing pressure. At ambient conditions, X-ray diffraction identified in $Li_3N$ a mixture of two phases: α-$Li_3N$ (P6/mmm) and metastable β-$Li_3N$ ($P6_3$/mmc); at about 0.5 GPa, α-$Li_3N$ fully transforms to β-$Li_3N$; a new phase γ-$Li_3N$ (Fm$\bar{3}$m) occurs near 40 GPa.[5,6] Normal ionic materials usually become metallic with increasing pressure, but $Li_3N$ is abnormal since the increasing pressure makes it into a much more strongly ionic state.[7] $LiN_3$ has been widely used in the industry as nitrogen sources, initial explosives and photographic materials.[8] Before 2013, $LiN_3$ was known in a single phase C2/m, and $LiN_3$ seemed so simple and well understood. However, several other phases of $LiN_3$ have been found using evolutionary crystal structure exploration methods coupled with first-principles calculations two years ago. At above 36 GPa, a hexagonal phase (P6/m) of $LiN_3$ with pseudo-benzene $N_6$ ring has been predicted by two research groups independently.[9,10] Some other phases appear as metastable: P$\bar{1}$ with a polyacetylene-like infinite linear nitrogen chain structure; C2/m and P$\bar{1}$ with puckered extended 2D decagonal and quasi-2D hexagonal nitrogen layers, respectively.[10] Above 375 GPa, Wang *et al* identified the phase of $P2_1$ which consists of zigzag N polymeric chains with $N_5$ ring sharing N-N pairs.[11] The band structures indicate that there are two metal-insulator transitions in $LiN_3$: first from insulator to metal at 36 GPa, and then from metal back to insulator at 375 GPa.

Besides nitrides (with $N^{3-}$ anion) and azides ($[N_3]^-$), in 2001, Kniep *et al* proved the existence of diazenides $[N_2]^{2-}$ by synthesizing $SrN_2$ and $BaN_2$ under high $N_2$ pressure.[12, 13] Since then, exploring new diazenides has been of constant interest. In 2010, alkali diazenides $Na_2N_2$ and $Li_2N_2$ (Pmmm) were predicted,[14] but then, a different structure of $Li_2N_2$ (Immm) was obtained under HP/HT conditions (9 GPa, 750 K) by decomposition of $LiN_3$.[15] This discrepancy encourages us to study $Li_2N_2$ under high pressure in detail.

Nitrogen can form many anionic species, e.g. $[N_2]^-$, $[N_2]^{3-}$ and $[N_5]^-$, which have just been obtained in molecular complexes.[16-19] We wonder if solid-state compounds with these anions in Li-N system could be synthesized under high pressure. Evolutionary algorithm USPEX has been widely used to predict new ground state structures in various systems without any experimental information, such as B-H, Xe-O, and Na-Cl.[20-22] The predicted counterintuitive compounds $NaCl_3$ and $Na_3Cl$ in the Na-Cl system have been confirmed by the experiment.[22] In this work, we have performed extensive structure searches on the Li-N system using variable-composition evolutionary algorithm USPEX, and indeed found many new stable compounds with very diverse and unusual crystal structures.

**Methods**

To search for stable compounds, the Li-N system was first explored using the variable-composition technique, as implemented in the USPEX code.[23-25] Evolutionary crystal structure predictions were performed in the pressure range from 0 to 100 GPa. Then we performed detailed fixed-composition evolutionary calculations the most promising compositions. All structure relaxations and electronic structure calculations were done using the Vienna Ab Initio Simulation Package (VASP) in the framework of density functional theory.[26] The Perdew-Burke-Ernzerhof generalized gradient approximation (PBE-GGA) was employed to treat the exchange-correlation energy,[27] and the all-electron projector augmented wave (PAW) potentials were used to describe the core-valence interactions.[28] The cut-off energy of 650 eV and Monkhorst-Pack k-point meshes for sampling the Brillouin zone with resolution $2\pi \times 0.04$ Å$^{-1}$ ensured that all the enthalpy calculations were well converged to better than 1 meV/atom. To ensure that the structures of predicted compounds in Li-N system are dynamically stable, phonon calculations were carried out using the Phonopy code.[29]

**Results and discussion**

We first studied the phase stabilities of Li-N system by calculating the enthalpy of formation (ΔH) for Li-N compounds in the pressure range from 0 to 100 GPa. The stability of compounds is determined by the thermodynamic convex hull construction. If the enthalpy of decomposition of a compound into any other compounds is positive, then the compound is stable, which is depicted on the convex hull. The convex hulls are shown in Figure 1 at selected pressures: 0, 20, 50, and 100 GPa. The various known phases of solid Li, $N_2$, $Li_3N$, $LiN_3$, and $Li_2N_2$ are reproduced readily in our evolutionary structure searches. Interestingly, five previously unreported compositions of Li-N system: $Li_{13}N$, $Li_5N$, $Li_3N_2$, $LiN_2$, and $LiN_5$ are found to be on the convex hull under ambient or high pressure in our calculations. The calculated phonon spectra confirmed that all compounds of five different compositions in Li-N are dynamically stable. Hence, we have found not only three new N-rich compounds, but also two new Li-rich compounds.

The main results are summarized as follows: (i) At ambient conditions (0 GPa), besides $Li_3N$ and $Li_2N_2$, $LiN_2$ with space group $P6_3/mmc$ is surprisingly stable. These three compositions are always stable in the pressure range from 0 to 100 GPa. (ii) However, the long-known LiN$_3$ is metastable below 48.7 GPa, which is in agreement with the known fact that it decomposes into N$_2$ and Li under external influences (heat, irradiation, etc) at 0 GPa. (iii) At 20 GPa, LiN$_5$ becomes stable, meanwhile Li$_{13}$N, Li$_3$N$_2$ and LiN$_3$ lie very close to (or nearly on) the convex hull. At 50 GPa, Li$_{13}$N, Li$_3$N$_2$ and LiN$_3$ are all stable, and Li$_5$N lies very close to the convex hull. At 100 GPa, Li$_5$N is stable, however, Li$_{13}$N and Li$_3$N$_2$ are becoming metastable although they both lie nearly on the convex hull.

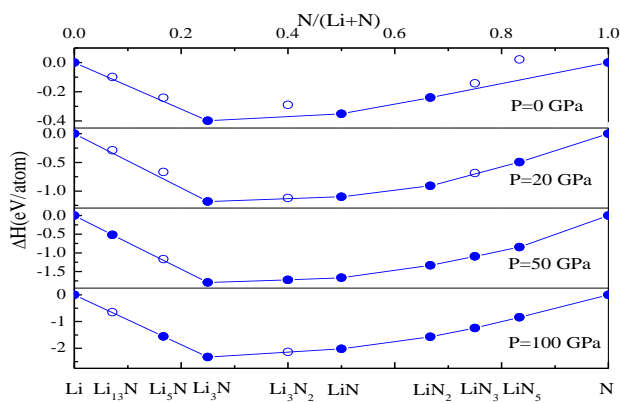

Figure 1. Enthalpies of formation (ΔH) of Li-N compounds from ground-state Li and N at 0, 20, 50, and 100 GPa.

The pressure-composition phase diagram of the Li-N system is depicted in Figure 2. For pure Li, with the increasing pressure, the bcc phase (Im$\bar{3}$m) transforms into fcc (Fm$\bar{3}$m), cI16 (I$\bar{4}$3d), Aba2-40, and Pbca phases in sequence, which is in accordance with previous experimental and theoretical data.[30-32] For pure N, the known Pa$\bar{3}$, P2$_1$/c, P4$_1$2$_1$2, and I2$_1$3 structures are reproduced in our searches and agree well with other theoretical predictions.[33,34] For Li$_3$N at ambient conditions, the stable phase is Pm$\bar{3}$m (not the experimental result P6/mmm, which we predict to be 22.32 meV/formula unit less stable at 0 GPa), and remains stable up to 0.2 GPa. The subsequent phases of Li$_3$N in our calculations are in agreement with the previous works.[5-7] For LiN$_3$, we also found a new metastable structure P$\bar{6}$2m in the pressure range 0-0.9 GPa. The Pm$\bar{3}$m phase of Li$_3$N and P$\bar{6}$2m of LiN$_3$ have not been found experimentally before, which may be because they are just stable in a rather small pressure range. For LiN (actually $Li_2N_2$), the obtained structure at 0 GPa is Pmmm, which is consistent with the theoretical result.[14] At 8.2 GPa, Pmmm phase of LiN transforms into the Immm structure, indicating that the experimental result obtained at around 9 GPa is also perfectly correct.[15] At 8.9 GPa, the Immm structure will lose its stability and the LiN phase Pnma becomes stable in the pressure range from 8.9 to at least 100 GPa.

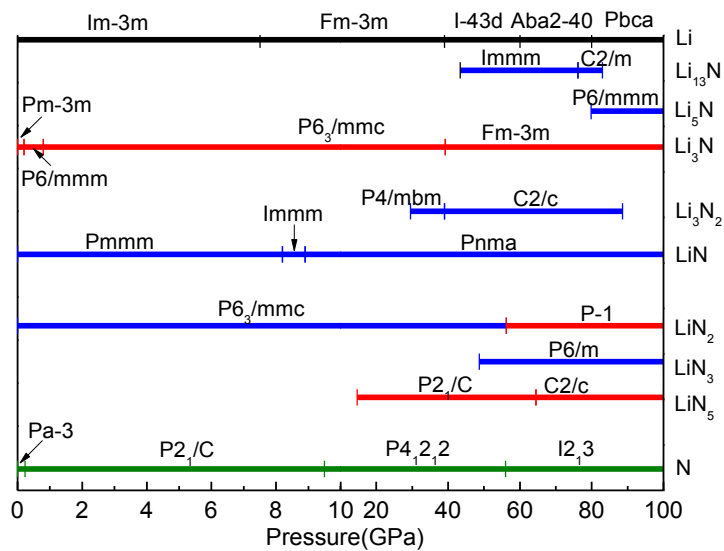

Figure 2. Pressure-composition phase diagram of Li-N system from 0 to 100 GPa. Except pure Li and N, the blue and red colors represent the metals and insulators characteristic of the Li-N compounds, respectively.

The phase transformations of five new compositions of Li-N system are as follows: (i) For $Li_{13}N$, the Immm structure is predicted to be stable from 43 to 76 GPa, following which the C2/m structure is stable up to 83 GPa. In fact, Immm and C2/m phases have nearly identical enthalpies (within 0.2 meV/atom), suggesting that $Li_{13}N$ can exist as a mixture of Immm and C2/m phases in its whole range of stability of this compound. (ii) $Li_5N$ has a single stable phase P6/mmm from 80 to at least 100 GPa. (iii) From 30 to 89 GPa, $Li_3N_2$ has two stable phases P4/mbm and C2/c. The pressure-induced structural transition from P4/mbm to C2/c occurs at about 39 GPa. (iv) Besides the P6$_3$/mmc structure, LiN$_2$ has another phase P$\bar{1}$, stable above 56 GPa, (v) For LiN$_5$, the P2$_1$/c structure becomes stable at 15 GPa and then transforms into C2/c phase when pressure is above 65 GPa. The C2/c structure is stable at least up to 100 GPa.

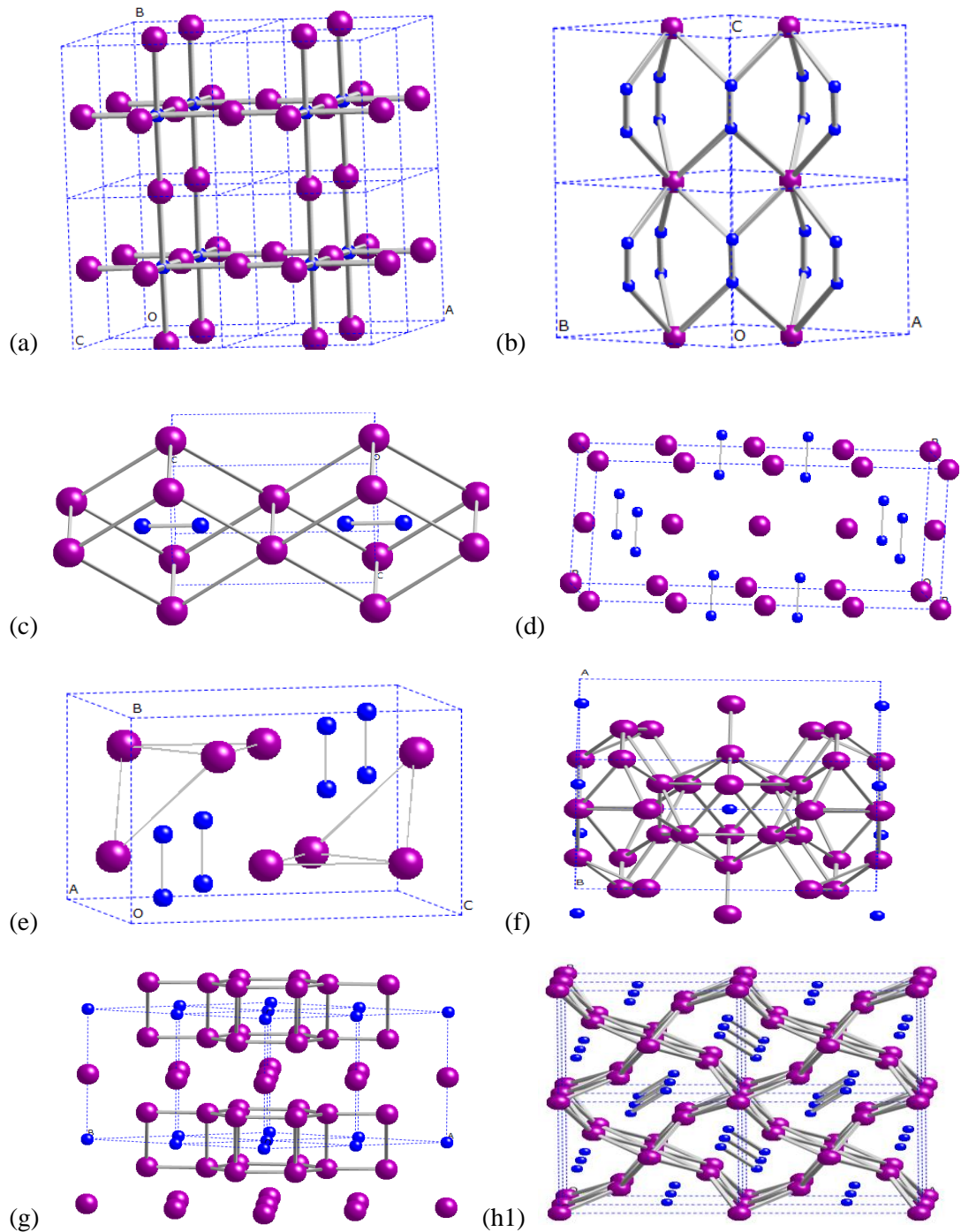

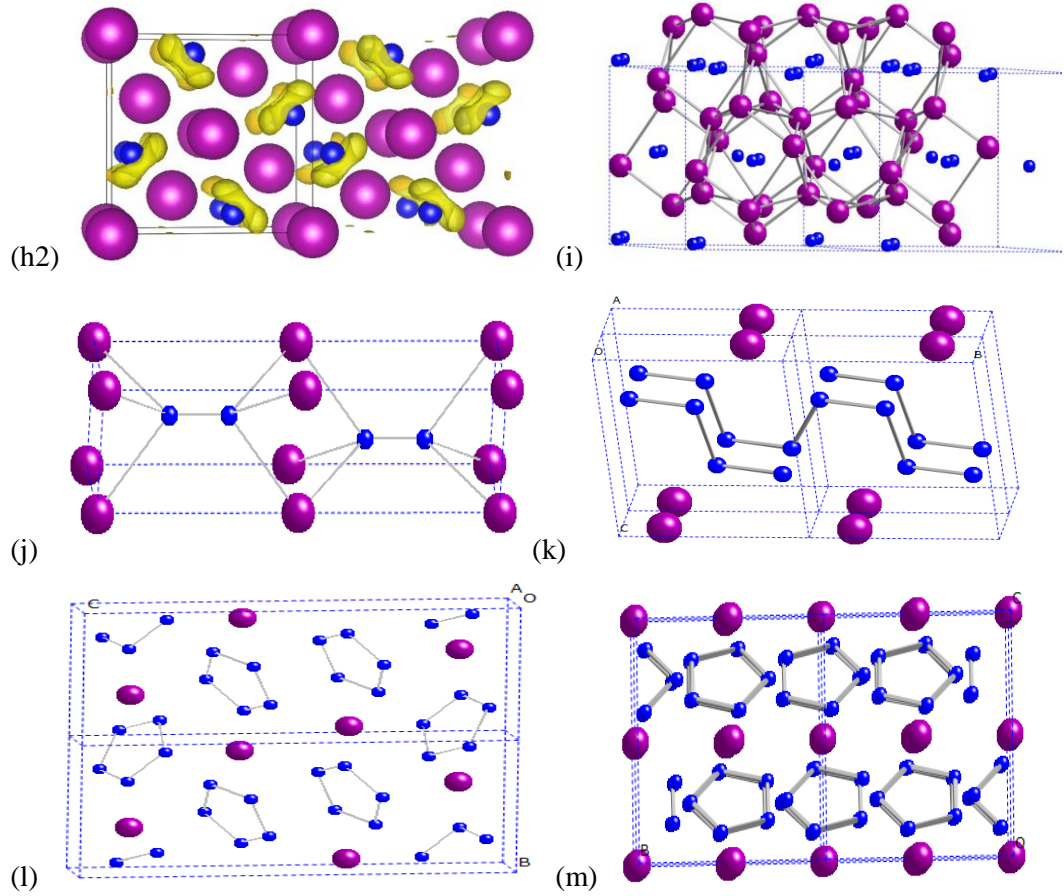

Figure 3. Crystal structures of Li-N compounds. (a) Pm$\bar{3}$m-Li$_3$N at 0 GPa, (b) P$\bar{6}$2m-LiN$_3$ at 0 GPa, (c) Pmmm-Li$_2$N$_2$ at 0 GPa, (d) Immm-Li$_2$N$_2$ at 10 GPa, (e) Pnma-Li$_2$N$_2$ at 10 GPa, (f) Immm-Li$_{13}$N at 50 GPa, (g) P6/mmm-Li$_5$N at 90 GPa, (h1) P4/mbm-Li$_3$N$_2$ at 30 GPa, (h2) ELF isosurfaces (ELF=0.85) of (h1), (i) C2/c-Li$_3$N$_2$ at 40 GPa, (j) P6$_3$/mmc-LiN$_2$ at 0 GPa, (k) P$\bar{1}$-LiN$_2$ at 60 GPa, (l) P2$_1$/c-LiN$_5$ at 50 GPa, (m) C2/c-LiN$_5$ at 80 GPa

The representative structures of above-mentioned Li-N compounds under ambient conditions and high pressure are presented in Figure 3. We first analyze the structures of Li$_3$N and LiN$_3$, and remind that the lengths of nitrogen-nitrogen bonds are 1.10 Å for the triple N≡N bond, 1.25 Å for the double N=N bond, and 1.45 Å for the single N-N bond. (i) For Li$_3$N, the calculated lattice parameters of P6/mmm, P6$_3$/mmc and Fm$\bar{3}$m are in agreement with experimental data within 0.5%. The Pm$\bar{3}$m structure is very simple, an anti-ReO$_3$-type structure made of corner-sharing NLi$_6$ octahedra (Figure 3a). Interestingly, across the phase transitions, the number of Li atoms surrounding each N atom increases from 6 for Pm$\bar{3}$m to 8 for P6/mmm, 11 for P6$_3$/mmc and 14 for Fm$\bar{3}$m. (ii) for LiN$_3$ at ambient conditions, C2/m structure consists of Li$^+$ cations and linear azide anions [N$_3$]$^-$.[10] As illustrated in Figure 3b, unlike the C2/m structure, P$\bar{6}$2m phase does have [N$_3$]$^-$ anions, but instead its unit cell contains two Li atoms and three N$_2$ groups with the N-N distance of 1.151 Å at 0 GPa, smaller than that in the azide-ion [N$_3$]$^-$ (1.184 Å), but larger than that in the gas-phase N$_2$ molecule (1.10 Å) and indicating a bond order between 2 and 3.

The Pmmm structure of Li$_2$N$_2$ consists of face-sharing Li$_8$ parallelepipeds (Figure 3c). The N$_2$ groups sits in the center of parallelepipeds, which can be viewed that there are six Li atoms connecting to each N atom of N$_2$ molecule and each of four Li atoms connects to both N atoms. The N-N bond length is 1.263 Å at 0 GPa, slightly larger than that of Na$_2$N$_2$ (1.24 Å)[14] and indicating a double N=N bond and ideal charge of the N$_2$ group equal to -2, which matches perfectly the formula Li$_2$N$_2$. Our calculated lattice constants of Immm structure (Figure 3d) are in good agreement with experimental results.[15] The predicted N-N bond length is 1.271 Å, slightly smaller than the experiment. Figure 3e presents the Pnma structure of Li$_2$N$_2$ at 10 GPa. Its unit cell contains four N$_2^{2-}$ groups and eight Li$^+$ ions. The N-N bond length is 1.269 Å.

For Li$_{13}$N, Immm and C2/m phases have similar structures. The Immm structure of Li$_{13}$N at 50 GPa is shown in Figure 3f. This structure is an interesting example of Li-N compounds which can be viewed as a combination of a single Li atom and a slightly distorted Li$_{12}$N icosahedron. The Li atoms of icosahedron include two different parts: the top and bottom each has a single Li atom, and the middle part consists of two opposite pentagon Li$_5$ rings. The N atom is enclosed in the center of the Li$_{12}$N icosahedron, which is analogous to Li$_{12}$Cs icosahedron in the Pnna structure of $Li_3Cs$ compound.[35] In $Li_3Cs$, the icosahedra share Li-Li edges with one another. However, the $Li_{12}N$ icosahedra are isolated and do not share Li atoms each other in $Li_{13}N$ compound. The Li-N bond lengths in the $Li_{12}N$ icosahedron are 1.934, 1.951 and 2.011 Å, i.e. nearly identical, and Li-Li distances are also nearly identical, ranging from 2.026 to 2.113 Å (maximum difference 4.3%, to compare with 22.3% in $Li_3Cs$[35]).

P6/mmm phase of $Li_5N$ has a layered structure, made of alternating layers of stoichiometry $Li_4N$ (here, N atoms are sandwiched between two Li-graphene sheets) and Li, see Figure 3g. Such unusual layered structures with alternation of "metallic" and "non-metallic" layers have been previously reported by some of us for the Na-Cl system (e.g., $Na_3Cl$, also confirmed experimentally[22]) and for the K-Cl system[36]. Bader analysis shows that $Li_5N$ at 90 GPa has charge configuration $[Li_4N]^{-0.68} Li^{+0.68}$, indicating that most of the valence electrons of Li layer transfer to $Li_4N$ sandwich layer.[37] Interestingly, the Bader charge of Li atom in upper Li-graphene sheet of $Li_4N$ sandwich layer is nearly neutral (+0.1e) and the charge of Li atom in bottom Li-graphene sheet is +0.74e.

As observed in Figure 3h1, the P4/mbm structure of $Li_3N_2$ consists of a three-dimensional network of Li atoms, which has open channels along z direction. This structure is very similar to the structure of the new compound $Mg_3O_2$ predicted by some of us recently,[38] except that in $Li_3N_2$ there is pairing of N atoms with the N-N distance of 1.353 Å at 30 GPa, indicating bond order between 1 and 2. Just like in P4/mbm-$Mg_3O_2$, we can clearly see columns of face-sharing body-centered cubes of metal atoms. The electron localization function (ELF) of $Li_3N_2$ (Figure 3h2) shows strong charge transfer from Li to N. However, unlike $Mg_3O_2$ which is an electride, there is no strong interstitial electron location in $Li_3N_2$. The Bader analysis also confirms the above result. The charges of P4/mbm-Li$_3$N$_2$ are +0.794e for one Li atom, +0.809e for the other two Li atoms, and -1.146e and -1.266e for two N atoms, respectively. The C2/c structure has a more complex three-dimensional network of lithium atoms with N$_2$ groups also sitting in its channels (Figure 3i), with the N-N distance of 1.391 Å at 40 GPa.

The P6$_3$/mmc structure of LiN$_2$ can be described as a NiAs-type structure, where anionic positions are occupied by the N$_2$ groups (Figure 3j). At zero pressure, the N-N distance is 1.173 Å, indicating a bond order between 2 and 3. The P$\bar{1}$ phase contains an infinite polyacetylene-like nitrogen chain (Figure 3k), similar to the metastable phase of LiN$_3$.[10] The N-N distances are 1.316, 1.320 and 1.333 Å at 60 GPa, suggesting bond order between 1 and 2. We can clearly see how pressure destroys molecular groups, favoring extended structures.

As observed in Figure 3l, the P2$_1$/c structure of LiN$_5$ consists of isolated Li atoms and N$_5$ rings which up to now were only detected in molecular complexes.[19] At 50 GPa, the N-N distances are 1.286, 1.291, 1.299, 1.303 and 1.305 Å, respectively. The higher-pressure C2/c phase also consists of isolated Li atoms and N$_5$ rings (Figure 3m). Unlike in P2$_1$/c, the N$_5$ ring here is an isosceles pentagon, with N-N distances of 1.277, 1.277, 1.301, 1.301, and 1.281 Å, respectively, at 80 GPa.

To obtain deeper insight into these new Li-N compounds, we calculated their band structures and density of states (DOS) at selected pressures. We found that all stable phases of Li$_{13}$N, Li$_5$N and Li$_3$N$_2$ at pressures are metallic. The Pm$\bar{3}$m of Li$_3$N is a semiconductor with the DFT band gap of 0.84 eV. Three phases of Li$_2$N$_2$ are also metallic, in agreement with experiment.[15] Interestingly, LiN$_2$ has a metal-insulator transition: P6$_3$/mmc is metallic at low pressure, but semiconducting in the high-pressure P$\bar{1}$ phase, with the band gap of 0.13 eV at 60 GPa. The new metastable P$\bar{6}$2m phase of LiN$_3$ is also metallic, hence the transitions of LiN$_3$ are exotic: from metallic to insulator to metallic to insulator. The P2$_1$/c and C2/c phases of LiN$_5$ are wide-gap insulators: e.g., the DFT band gap of the C2/c phase at 80 GPa is 2.1856 eV.

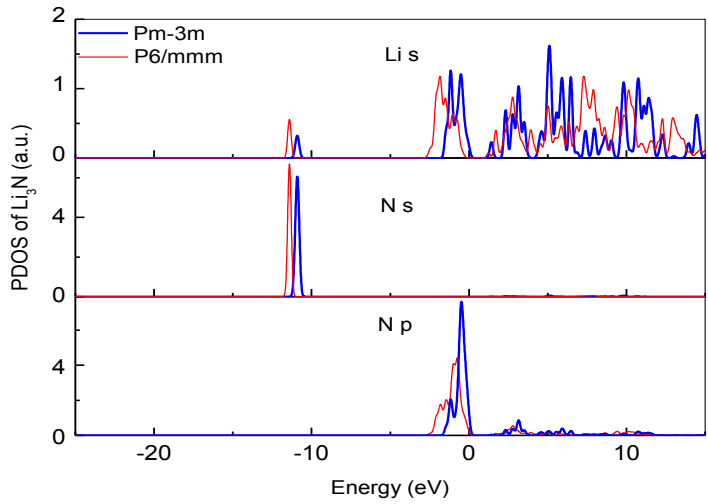

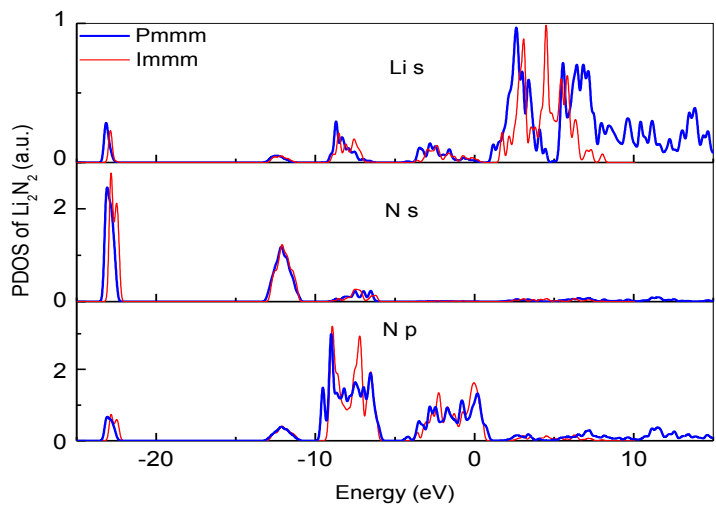

(b)

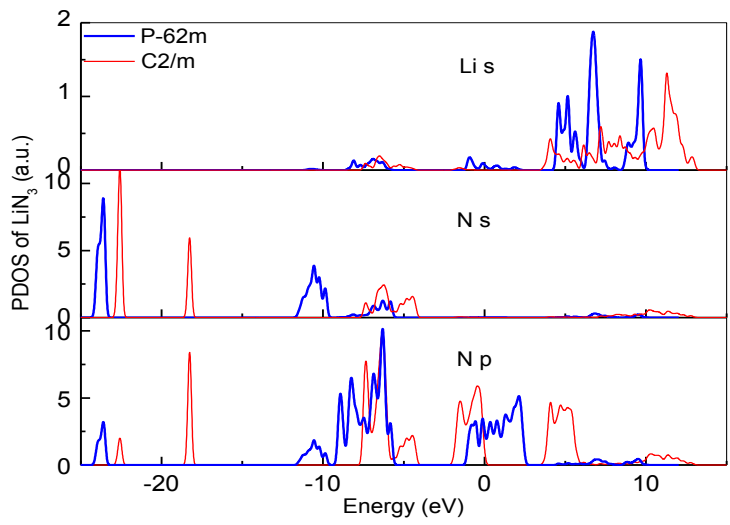

(c)

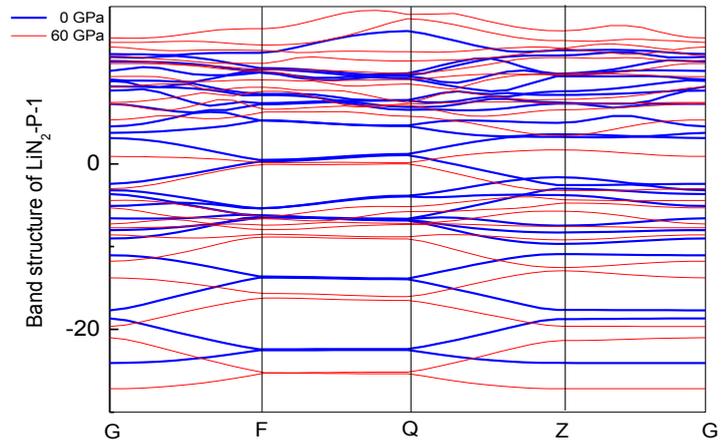

(d)

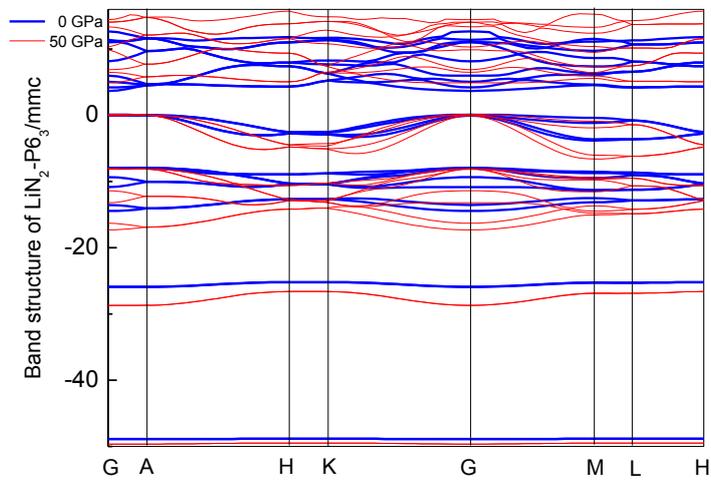

(e)

Figure 4. Electronic structures of selected Li-N compounds. (a)-(c) PDOSs of $Li_3N$, $Li_2N_2$ and $LiN_3$ at 0 GPa. (d)-(e) Band structures of $P\bar{1}$-$LiN_2$ and $P6_3/mmc$-$LiN_2$ at two different pressures.

Some electronic structures are shown in Figure 4. As is seen from Figure 4 (a)-(c), the PDOSs of two different phases of $Li_3N$ (or $Li_2N_2$ or $LiN_3$) at 0 GPa are obviously different. The new-found phases $Pm\bar{3}m$-$Li_3N$, $Pmmm$-$Li_2N_2$ and $P\bar{6}2m$-$LiN_3$ have one character in common: the states near the Fermi level come mostly from Li s and N p orbitals. Figure 4 (d)-(e) show the band structures of $P\bar{1}$-$LiN_2$ and $P6_3/mmc$-$LiN_2$ at different pressures, respectively. The band structures at different pressures for the same phase are similar. When pressure increases, the band structure is more dispersive and the bandwidths also increase: both the conduction and valence bands broaden, and conduction band tends to shift upwards in energy. These changes can lead to both metallization and demetallization: $P\bar{1}$-$LiN_2$ is metallic at 0 GPa, whereas it becomes semiconductor with the gap of 0.13 eV at 60 GPa.

**Conclusions**

In conclusion, we have predicted a number of new Li-N compounds using *ab initio* evolutionary structure search. Other than the well-known compositions $Li_3N$, $Li_2N_2$ and $LiN_3$, we found five novel compositions which should be experimentally synthesizable under pressure, including $Li_{13}N$, $Li_5N$, $Li_3N_2$, $LiN_2$, and $LiN_5$. Notably, the N-N bonding patterns evolve from isolated N ions to $N_2$ dumbbells, to linear $N_3$ groups, infinite nitrogen chains, $N_5$ rings with increasing N content. Interestingly, at ambient conditions, we also identified a new stable phase $Pm\bar{3}m$ of $Li_3N$ and a new metastable phase $P\bar{6}2m$ of $LiN_3$. This work resolves previous discrepancy on stable phases of $Li_2N_2$ and provides the basis for the future experimental investigations of the Li-N system.

**Acknowledgments**

Y. Shen thanks Dongxu Li, Xiao Dong, Xiangfeng Zhou, Qianku Hu, Shengnan Wang and Haiyang Niu for valuable discussions. The research is supported by National Natural Science Foundation of China (No.11204053 and No.11074059) and the China Postdoctoral Science Foundation (No. 2013M531028). A. R. O. thanks DARPA (No.W31P4Q1210008 and No.W31P4Q1310005) and the Government of Russian Federation (No.14.A12.31.0003) for financial support.